\documentclass[graybox]{svmult}
\usepackage{color,soul}

\usepackage{type1cm}        
\usepackage{makeidx}         
\usepackage{graphicx}        
\usepackage{multicol}        
\usepackage[bottom]{footmisc}

\usepackage{newtxtext}       %
\usepackage[varvw]{newtxmath}       

\makeindex             

\begin{document}

\title*{Computerized Modeling of Electrophysiology and Pathoelectrophysiology of the Atria\\ 
–- How Much Detail is Needed?}
\titlerunning{Computerized Modeling of the Atria}
\author{Olaf Dössel\orcidID{0000-0003-4551-7970} and\\ Axel Loewe\orcidID{0000-0002-2487-4744}}
\institute{Axel Loewe and Olaf Dössel\at Institute of Biomedical Engineering, Karlsruhe Institute of Technology (KIT), Kaiserstr. 12, 76131 Karlsruhe, Germany, \email{publications@ibt.kit.edu}}
%
%
\maketitle

\abstract{
This review focuses on the computerized modeling of the electrophysiology of the human atria, emphasizing the simulation of common arrhythmias such as atrial flutter (AFlut) and atrial fibrillation (AFib). Which components of the model are necessary to accurately model arrhythmogenic tissue modifications, including remodeling, cardiomyopathy, and fibrosis, to ensure reliable simulations? The central question explored is the level of detail required for trustworthy simulations for a specific context of use. The review discusses the balance between model complexity and computational efficiency, highlighting the risks of oversimplification and excessive detail. It covers various aspects of atrial modeling, from cellular to whole atria levels, including the influence of atrial geometry, fiber direction, anisotropy, and wall thickness on simulation outcomes. The article also examines the impact of different modeling approaches, such as volumetric 3D models, bilayer models, and single surface models, on the realism of simulations. In addition, it reviews the latest advances in the modeling of fibrotic tissue and the verification and validation of atrial models. The intended use of these models in planning and optimization of atrial ablation strategies is discussed, with a focus on personalized modeling for individual patients and cohort-based approaches for broader applications. The review concludes by emphasizing the importance of integrating experimental data and clinical validation to enhance the utility of computerized atrial models to improve patient outcomes.}

\section{Introduction}
\label{sec:intro}
The topic of this article is computerized modeling of the electrophysiology of the human atria, including the simulation of frequent arrhythmias, such as atrial flutter (AFlut) and atrial fibrillation (AFib). This also demands modeling of arrhythmogenic tissue modifications, such as disease-induced remodeling and atrial cardiomyopathy, including fibrosis~\cite{Goette-2024-ID19552,Schotten-2024-ID19723}. We will focus on the question: `How much detail is really needed'? This seems to be a general principle: the more we look into the details of biological systems, the more complex they become. Generic models for specific species divide into different and more complex systems when we move to personalized models. But too many details pose risks and challenges to models: not only do calculation times increase but we might lose oversight and leave the regime of identifiable parameters for a given set of observations. As John von Neumann expressed it: ``With four parameters I can fit an elephant, and with five I can make him wiggle his trunk''~\cite{Dyson-2004-ID19724}. However, neglecting important details will lead to unrealistic results, predictions will not be reliable, and the model might fail to generalize and extrapolate beyond the data used to build it.
The quality of any computerized modeling in medicine and in particular the answer to the question of the degree of detail that is needed can only be answered if an 'intended purpose' or 'context of use' of the model is specified~\cite{Viceconti-2024-ID19278,Viceconti-2021-ID15999}. In this article, the intended use is the planning and optimization of atrial ablation to stop reentrant arrhythmia and prevent recurrence. This includes both personalized modeling of the individual patient (digital twin) and modeling of cohorts with specific characteristics so that rules can be derived for optimal ablation strategies in subgroups with specific characteristics.

\begin{figure}[ht]
\includegraphics[width=\textwidth]{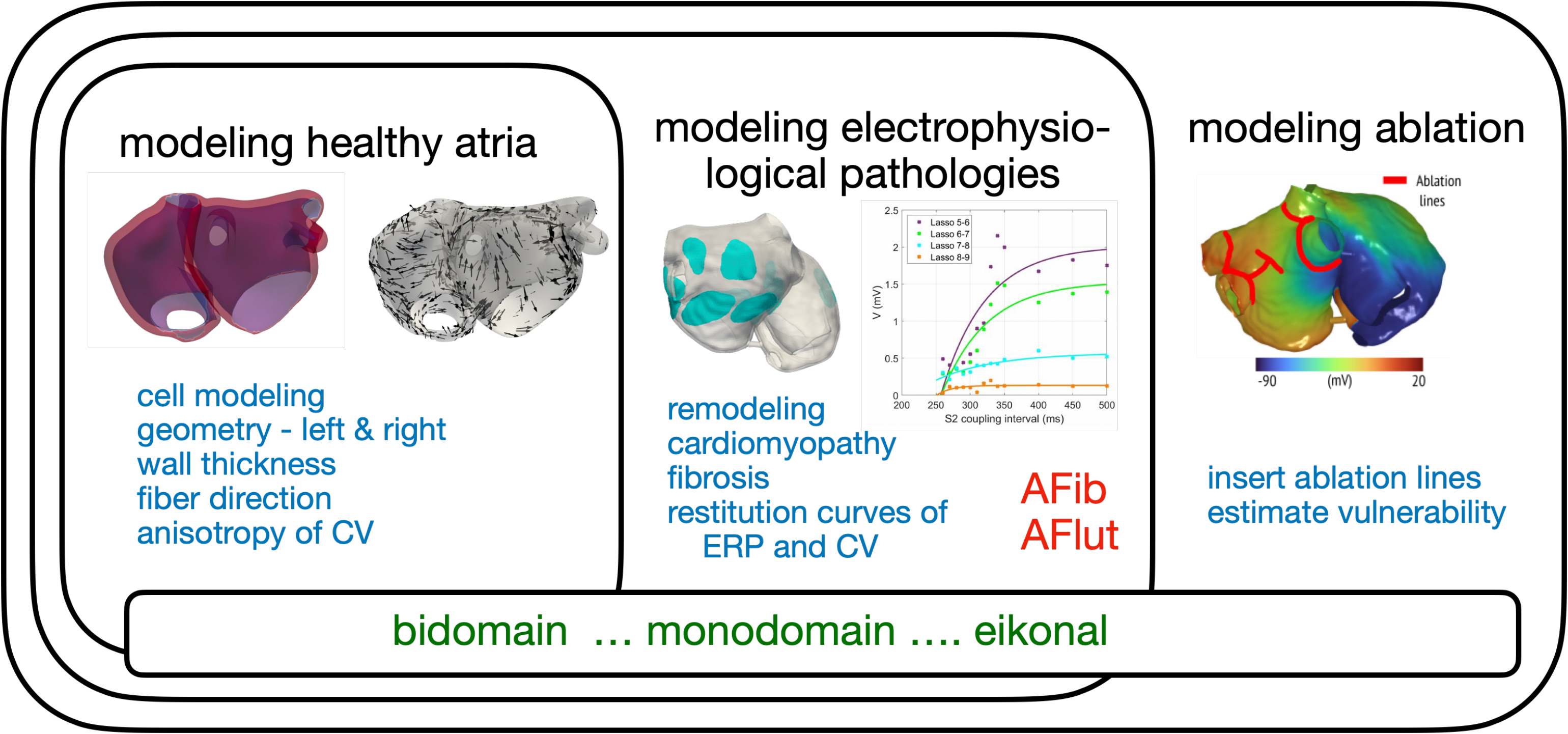}
\caption{Aspects of computerized modeling of electrophysiology and pathoelectrophysiology of the atria with planning of successful ablation as the intended use.}
\label{fig:Aspects}       
\end{figure}

This article moves along the biological levels of integration from cell modeling to modeling the whole atria. It covers aspects of heterogeneity, and modeling of diseased, remodelled and fibrotic tissue. It includes the influence of atrial geometry, left and right atria, wall thickness, fiber direction and anisotropy on simulation results. Figure~\ref{fig:Aspects} gives an outline of these aspects of computerized modeling of the atria.

This article reports on progress since a previous review~\cite{doessel12}. Review articles with different scope and focus but partial overlap were published lately~\cite{Kwon-2021-ID19403,Heijman-2021-ID18926,Loewe-2022-ID18372,Trayanova-2024-ID19465,Jaffery-2024-ID19444,Cluitmans-2024-ID19340}. Simulation and exploitation of ECG signals are not in the focus of this article and reviewed elsewhere~\cite{Dossel-2021-ID16522}.

\section{Cellular Modeling}
\label{sec:cellular}
11 years ago, we published an article comparing various atrial cell models~\cite{wilhelms13}: the Courtemanche et al.\ model, the Grandi et al.\ model, the Maleckar et al.\ model and the Koivoumäki et al.\ model, and we observed small differences in the slope of the action potential but major differences in Ca$^{2+}$ handling, which will have strong influence on tension development. Meanwhile, there are several new cell models like e.g. Mazhar et al.~\cite{Mazhar-2023-ID18743}, but we do not intend to compare them all in this article. The Courtemanche model is the one that is still used most often.
The conductivity of the ion channels of one type ($g_{i,\mathrm{max}}$) are not equal in all cells in a tissue  but show a distribution~\cite{Dasi-2024-ID19471}. These differences are smeared out to a large extent, when the cells are coupled via electrotonic coupling (e.g. via gap junctions).
Cell models need modifications to adapt them to the properties of specific areas in the atria like e.g. pectinate muscles, the crista terminalis, the interatrial connections (e.g. Bachmann's bundle)~\cite{seemann06,krueger13b} and can be adapted to reflect gene mutations~\cite{loewe14,loewe15e} or the effect of drugs~\cite{loewe14b,Dasi-2023-ID18696,Wiedmann-2020-ID14070}. 

Mechanoelectrical coupling will lead to modifications in cellular electrophysiology due to mechanical stress in atrial tissue. Since contraction of the atria during atrial systole is quite small, this effect might be neglected. But ventricular contraction is leading to significant mechanical stress in the atria. In computer simulations Appel et al. have demonstrated that this can have an effect on effective refractory period and might induce ectopic beats ~\cite{Appel-2024-ID19734}.

\subsection{Electrophysiological Remodeling due to AFib}
\label{sec:remod}
Cell models of the atria have to be modified to represent remodeling processes in response to AFib~\cite{krueger13}. Experimentally reported results on changes of ion channel conductance and kinetics vary but for most channels a clear tendency can be condensed leading to an overall shorter action potential in cells that experienced persistent AFib conditions~\cite{loewe14a}.

\subsection{Sinus Node Modeling}
\label{sec:sinus}
Models of the sinus node have become more and more sophisticated. Obviously, we need cell models that depolarize periodically in an autonomous manner driven by an interplay of the “funny” ion current I\textsubscript{f} , the Ca$^{2+}$ homeostasis and the Na$^+$/ Ca$^{2+}$ exchanger NCX (`calcium clock')~\cite{Kohajda-2020-ID13995}. In addition, these pacemaker cells have to be embedded into the atrium in a specific way allowing the atrial myocardial cells to capture the excitation without putting too much electrical load on the sinus node cells (`pace and drive'), for example by conductivity gradients and non-conducting structures to protect the sinus node cells from current drain~\cite{Amsaleg-2022-ID18319,Ricci-2024-ID19546}. Modeling the sinus node properly is a prerequisite for understanding sinus node arrhythmias like bradycardia~\cite{Loewe-2019-ID12801}, for research on the influence of the autonomous system~\cite{Linder-unknown-ID19579} and of the effect of drugs on the heart rate. Sinus node modeling also becomes important if variations in P-wave morphology are of interest~\cite{loewe16e} or if the interaction between sinus beats and reentrant/ectopic activation is relevant for an arrhythmia. In many cases the effect of the sinus node can be represented by specifically placed and timed stimuli without explicit pacemaker cell models.

\section{Anatomical Modeling}
\label{sec:geom}
Obviously, the geometry of the atria varies between patients, and this will have strong influence on initiation and termination of AFlut and AFib. Large atria are known to be a promoter of AFib, but also eccentricity may play a role~\cite{Jia-2021-ID17070}. Nagel et al.\ published a statistical shape model of the geometry of the human atria~\cite{Nagel-2022-ID17346}. This shape model is based on 47 thoroughly segmented atria and generalizes well to unseen geometries with 95\% of the total shape variance being covered by its first 24 eigenvectors. Such a shape model can be used to create cohorts with thousands of different geometries of human atria reflecting the real shape distribution in the population used to create the shape model. It can also be used to register individual atrial models to a joint reference geometry~\cite{Nairn-2023-ID18842}. 

Most often, left and right atrial appendages are neglected in computer modeling of atrial electrophysiology. Thus, it is not precisely known, to what extent they have to be included.

Roney at al.\ presented a universal atrial coordinate system that allows for comparison of regional tissue characteristics across patients~\cite{Roney-2019-ID14879}. Corrado et al.\ investigated the impact of shape uncertainty on predicted atrial arrhythmias~\cite{Corrado-2023-ID18413}. They report that the vulnerability of an atrium cannot be predicted by transforming to a standard geometry with known vulnerability.

Standardized regions of the atria and rules to define them are needed if e.g. in a clinical study the average number of rotors in different regions of the atria is determined. If there is no consensus about how to define these regions, we cannot compare these clinical studies. Recently we published an article about the standardized and automatic regionalization of the atria~\cite{Goetz-2024-ID19745}.

\subsection{Fiber Direction}
\label{sec:fiber}
Ho has published her findings about the preferential myocyte orientation, also referred to as `fiber direction', in the atrial walls based on explanted hearts~\cite{ho09}. Submillimeter diffusion tensor MRI delivered more details~\cite{Pashakhanloo-2016-ID17330}. The fiber direction is much more complex than in the ventricles. In many regions, the fiber directions in the endo- and epicardial walls are different. The crista terminalis and the pectinate muscles have a distinct longitudinal fiber direction. We published a rule-based method to define a fiber direction that comes very near to the anatomical atlas~\cite{krueger11a,wachter15,Azzolin-2023-ID17490}. 
Roney et al.\ used very high resolution diffusion tensor magnetic resonance imaging (DT-MRI) to construct a human atrial fiber atlas~\cite{Roney-2020-ID14701}. DT-MRI could in principle deliver the fiber direction of an individual patient. But this might not be realistic for clinical applications, since the atrial wall in some areas is only 1\,mm to 2\,mm thick and the voxel size of MRI is about 1\,mm$^3$. A fiber difference between endocardium and epicardium can currently not be resolved with clinical MRI in vivo.

\subsection{Wall Thickness}
\label{sec:wall}
Azzolin et al.\ demonstrated that the wall thickness is another important aspect~\cite{Azzolin-0000-ID14715,kharche15,whitaker16,Wang-2019-ID12885}. A literature survey showed that the thickness of the atrial wall typically ranges from 1\,mm to 4\,mm depending on the region. By studying meandering of long living rotors, Azzolin recognized that they preferably anchor in areas of large gradients in wall thickness. We assume that locations of high thickness gradients are also prone to unidirectional block and therefore important when studying initiation/inducibility of reentry.

\subsection{Biatrial vs. Left Atrial Models}
\label{sec:biatrial}
Martínez Díaz et al.\ lately published the results of a study on the question: do we need the right atrium for good estimation of AFib vulnerability?~\cite{MartinezDiaz-2024-ID19306} The background of this question is the following: If we build our model on EP data, we usually only have the geometry of the left atrium. If we use MRI data, we do also have data about the geometry of the right atrium, but it is rarely used in the literature. In the study of Martínez Díaz et al., in 8 clinical cases, three degrees of fibrosis were introduced: healthy, moderate and severe. Then, vulnerability was assessed as the ratio of AFib inducing points divided by the number of stimulation test points. The results differed between the patients. In most of the cases, the vulnerability ratio increased with increasing degree of fibrosis. But not always: there is one case where it practically was not changing and one case where it was even the other way around. To answer the initial question: the overall picture is clear; the vulnerability ratio is higher in biatrial models in comparison to monoatrial models. Incorporating the right atrium in patient-specific computational models unmasked potential inducing points in the LA. Because virtual ablation strategies for AF rely on noninducibility criteria, performing biatrial simulations is advisable.

\subsection{Full 3D Model or Bilayer Model or Single Surface Model?}
\label{sec:volumetric}
An important aspect of the geometrical model is the question whether we need a full volumetric 3D model of the atrial walls, whether a bilayer structure~\cite{labarthe14} is sufficient or whether even a single surface, folded into the 3D space is good enough. A 3D model should at least include one layer of e.g. tetrahedra on each side of the atrial wall if the discrete fiber layers have to be considered. The average side length of the tetrahedra should therefore not be longer than 0.3\,mm. This results in millions of nodes in the computational mesh. 

The bilayer model stands for the epicardial and the endocardial side of the atrial wall. Thus, it can consider different fiber directions of the endo- and the epicardial side. It can even consider endo-epi-dissociation, which is recognized as an important substrate for AFib~\cite{degroot16}. Single layers, folded into 3D space, are even more simple and are very popular because of the fast calculation time. Since they cannot take the anisotropy of conduction across the atrial wall into account properly, the question whether they are suitable for in silico AFib vulnerability assessment remains open.

\section{Mathematical Modeling of Excitation Propagation}
\label{sec:MathMod}
The bidomain equations are the basis for most tissue level models of cardiac electrophysiology and are covered well in literature, where also the standard naming of the parameters is explained~\cite{vigmond02}: 

\begin{equation}
-\nabla \cdot\left(\left(\sigma_\mathrm{i}+\sigma_\mathrm{e}\right) \nabla \Phi_\mathrm{e}\right)=\nabla \cdot\left(\sigma_\mathrm{i} \nabla V_\mathrm{m}\right)~, 
\end{equation}
\begin{equation}
\nabla \cdot\left(\sigma_\mathrm{i} \nabla V_\mathrm{m}\right)+\nabla \cdot\left(\sigma_\mathrm{i} \nabla \Phi_\mathrm{e}\right)=\beta\left(C_\mathrm{m} \frac{\partial V_\mathrm{m}}{\partial t}+I_\mathrm{ion}-I_\mathrm{s}\right)~.
\end{equation}
The bidomain equations are one of the most detailed computer model for the spread of depolarization that we have today. But there are some approaches to go even beyond that level of detail~\cite{Tveito-2021-ID18922,Steyer-2023-ID18901}. Solving the bidomain equations is computationally expensive: typically several hours for one single heartbeat. The monodomain model reduces computational cost by assuming a linear relation between intracellular and extracellular conductivity tensors. They can remain anisotropic but point in the same direction. Further simplifying the model by assuming isotropic conductivities will not lead to realistic results~\cite{Fastl-2018-ID13241}. 

A next step of simplification is the eikonal equation: 

\begin{equation}
\sqrt{\nabla T(x)^{\top} \cdot M \cdot \nabla T(x)}=1~,    
\label{eq:eikonal}
\end{equation}
\begin{equation}
V_\textrm{m}(x, t)=U(x, t-T(x))~.    
\end{equation}

It only yields activation times and neglects effects of wavefront curvature and bath-loading at the surface of the tissue. $M$ in equation~\ref{eq:eikonal} can be a tensor, meaning that an anisotropic conduction velocity can be taken into account. There are attempts to include repolarization~\cite{Espinosa-2024-ID19535,neic17} and curvature of tissue~\cite{Skupien-2022-ID18179} (see chapter~\ref{sec:eikonal} about eikonal modeling of AFib for details). If we want to simulate electrograms (EGMs) or ECGs, we have to couple the eikonal equation with an assumption about the action potential, which starts in the moment the wave arrives at that point in space. The reaction eikonal model published by Neic et al.\ is a major improvement in this respect~\cite{neic17}. Here, the action potential is not a fixed precomputed curve, but it is calculated dynamically allowing for adaptation to the local situation.
The eikonal equation can be solved with the fast marching algorithm or the fast iterative method speeding up calculation significantly. On top of that, the eikonal equation can deliver quite accurate results even with a much coarser mesh as compared to bidomain and monodomain equations. The activation times are interpolated faithfully in-between the nodes. All this results in a speed up factor of several hundred~\cite{Espinosa-2024-ID19535,neic17}.

Nagel et al.\ compared the following propagation models: bidomain, monodomain, eikonal and reaction eikonal~\cite{Nagel-2022-Comparisonofpropag}. Conductivity tensors suggested by Clerc~\cite{Clerc-1976-ID12471} and those proposed by Roberts and Scher~\cite{roberts82} were used. The forward calculation of body surface potentials was carried out with the finite element method, the boundary element method and the infinite volume conductor method. Propagation models were compared in healthy atria and in AFib-remodeled atria with fibrosis in terms of action potentials, local activation times (LATs) and voltages measured on the body surface, i.e. the ECG. The largest differences in local activation times were due to different values for the conductivities. Comparing all simulation models with Clerc conductivities yielded nearly identical results. Thus, the very fast eikonal model together with the boundary element method was deemed enough for most applications in sinus rhythm. Reducing uncertainty of conductivity tensors both in healthy and in fibrotic tissue appears as important. The infinite volume conductor model failed to determine a precise ECG in particular in the precordial leads.

\subsection{Anisotropy and Restitution of Conduction Velocity}
\label{sec:anisotropy}
This is another level of complication: anisotropy of conduction velocity (CV) and CV restitution, i.e. rate dependence. Anisotropy of CV means that the conduction velocity depends on the actual direction of the depolarization wave with respect to the local fiber direction – and this angle can obviously be changing during AFlut and AFib. Scalar values of CV determined during sinus rhythm will not be valid during AFib~\cite{Coveney-2022-ID18047}. The restitution  of CV shows to what extent the CV (better: the effective CV in small tissue patches) is reduced if the time between two depolarization waves becomes small and smaller – which definitely happens during AFlut and AFib. Nothstein et al.\ carried out a clinical study in which they measured both anisotropy and restitution curves of CV~\cite{Nothstein-2021-ID16334}. They used the classical Lasso catheter and stimulated the atrial tissue of patients locally from 3 directions – one after the other - and with different S1/S2 coupling intervals. Measurements were performed in different atrial locations: some inconspicuous, presumably healthy and some clearly fibrotic with low electrogram amplitudes (`voltage'). Both voltage and CV were strongly lower in fibrotic areas with a steep decrease for coupling intervals shorter than ca. 350\,ms. This short coupling interval is typical for AFib. Thus, anisotropy of CV and the corresponding restitution curve have to be considered if we want to create a good computer model of AFib~\cite{Lubrecht-2021-ID15889,RuizHerrera-2022-ID18046,Nothstein-2021-ID16334,Becker-2024-ID19748}.

\subsection{Modeling Fibrotic Tissue}
\label{sec:fibrosis}
Fibrotic tissue can be both a cause and a consequence of atrial fibrillation. Fibrotic tissue contains collagen - interstitial and/or replacement - and is characterized by activated fibroblasts~\cite{Goette-2024-ID19600, Schotten-2024-ID19723, Sanchez-2022-908069}. Collagen cannot depolarize and cannot actively propagate the depolarization wave. With different degrees of fibrosis, the volume fraction of these non-myocyte areas becomes larger. Ways of incorporating fibrosis into computer models of cardiac electrophysiology are reviewed in~\cite{Sanchez-2022-908069}. Roney et al.\ compared different methods to model fibrosis~\cite{Roney-2016-ID12208} and found that the specific choice of model affects rotor dynamics and simulated electrograms. Keller et al.\ simulated cases of 20\% to 60\% replacement fibrosis and observed the spread of depolarization and the corresponding EGM~\cite{keller13}. Below 30\%, hardly any effect was visible on electrophysiology in terms of global depolarization patterns. Above ca. 60\%, the tissue becomes a so-called block: the depolarization wave cannot propagate through large fibrotic areas anymore in this model. 
Vigmond et al.\ introduced the concept of “percolation” into the field of modeling fibrotic tissue yielding very realistic simulated fragmented and fractionated EGMs~\cite{vigmond16}. Keller and Vigmond come to similar conclusions: above fibrotic tissue, the voltage becomes smaller with increasing degree of fibrosis while the effective, macroscopic conduction velocity becomes slower. In addition, the electrogram becomes more and more fragmented and fractionated. This is due to the tortuous zig-zag path of the depolarization wave leading to frequent orientation changes of the wavefront. Thus, we observe a strongly nonlinear effect on arrhythmogenicity as a function of the volume fraction of fibrosis: the most dangerous degree of fibrosis is between 30\% and 60\%. A segmentation of LGE-MRI into only two classes, fibrotic and non-fibrotic, is a dangerous oversimplification. In the chapter about verification, the simulation of EGMs above fibrotic tissue is further detailed. 

\subsection{Eikonal Model for Simulation of AFib}
\label{sec:eikonal}
The standard eikonal model can simulate just one activation sequence. Thus, it can be used for sinus rhythm and single cycles of AFlut. A multi-front eikonal model was suggested, that also allows for simulation of reentrant AFlut~\cite{pernod11,Loewe-2019-ID12386}. The simulation of AFib demands for reentrant depolarization waves.
Barrios et al.\ pushed the eikonal model further in this respect by proposing the “DREAM”: Diffusion Reaction Eikonal Alternant Model~\cite{AlbertoBarriosEspinosa-2022-ID18494,Espinosa-2024-ID19535}. The concept behind the reaction eikonal model~\cite{neic17} is extended by calculating repolarization times and making nodes available again for activation after repolarization. Alternative models for scar-related reentry and AFib were proposed as well~\cite{Gander-2024-ID19655,Gander-2023-ID19277,Campos-2021-ID16613}.

\section{Verification and  Validation of Computer Modeling of the Atria}
\label{sec:verification}
Validation answers the question whether a model correctly reproduces real depolarization patterns, activation times, EGMs and ECGs of the human atria, i.e. whether a model faithfully represents the underlying physiological system and is, thus, adequate for guiding ablation, which is the ``intended use'' in this article (see section~\ref{sec:intro},\cite{Galappaththige-2022-ID18204,Viceconti-2024-ID19278,Pathmanathan-2024-ID19555}). A prerequisite for this is technical verification in terms of whether a given simulation code provides a faithful representation of the solution of the governing equations for a specific numerical method~\cite{niederer11,Lindner-2022-ID18393,Pezzuto-2016-ID12466}. Verification and validation of models is often forgotten in the field of computer modeling of the heart. We find many publications for example with beautiful movies of mathematical rotors and how one can stop them, but the authors do not make any attempt to calculate measurable signals to compare with real signals from real patients. While prospective validation can be difficult, a lack of clinical electrical signals from the heart is not the problem in this case: thousands of ECGs and EGMs are acquired every day. We suggest that all computer modeling studies of the heart should systematically compare the simulated electric signals with measured signals, e.g. EGMs and ECGs. In addition, uncertainty quantification should be performed systematically to assess the effect of parameter uncertainty in the model~\cite{mirams16,Winkler-2022-ID18394}.

\subsection{Comparing Simulated and Measured Electrograms (EGMs)}
\label{sec:egms}
Comparing with EGMs is straightforward. One does not need a volume conductor model of the thorax that might introduce additional uncertainties. One can compare clinically important features of these signals like activation times, voltage, fractionation, activity duration and conduction velocity~\cite{keller13,vigmond16,Sanchez-2022-908069}.
There are several multichannel catheters for electrophysiology (EP) available (e.g. the Lasso, the PentaRay and the HD grid, mapping systems were critically reviewed by Groot et al.~\cite{deGroot-2021-ID17213}). By moving them inside the atrium, it is possible to pick up thousands of EGMs in a few minutes. Maps of diagnostically relevant properties can be visualized, e.g. voltage maps and maps of local activation time (LAT).

By using a simulation package like e.g. openCARP~\cite{Plank-2021-ID15953}, EGMs above healthy tissue can be simulated, and they show good similarity to measured EGMs~\cite{Steyer-2023-ID18649,Sanchez-2022-908069}. Simulating signals above fibrotic tissue is more involved~\cite{Sanchez-2022-908069}. Sánchez et al.\ simulated hundreds of EGMs with different degrees of fibrosis and different transmural extents~\cite{Sanchez-2021-ID16474}. The objective was to train a neural network with these synthetic data and find a quantitative estimate of the degree and the transmural extent of fibrosis from the EGMs. These estimates are important for cardiologists and for a trustworthy personalization of a computer model of a patient’s atria.

The bad news is that simulations clearly show that endocardial measurements with catheters can only show what's going on on the endocardial side. The epicardial side can only be seen as a small contribution to the signal. It is practically invisible if there is a strong signal from the endocardial side. The hypothesis that larger electrodes can better see the epicardial side is not true. If one increases the diameter of the electrode to 2\,mm, an epicardial layer of fibrosis will still only reduce the signal by a few percent~\cite{Nairn-2020-ID15709}. If the endocardial side is healthy, it will clearly dominate the EGM. Exclusive epicardial fibrosis can hardly be seen with endocardial electrodes. But it also becomes clear that an exclusively epicardial fibrosis does not influence the global spread of depolarization at all. Is it not relevant for modeling AFib? It is, because as soon as the effective CV on the epicardial side becomes extremely slow, the delayed wavefront from the epicardial side can find excitable tissue on the endocardial side leading to transmural reentries in case of impaired coupling between endocardial and epicardial myocardium.

\section{Simulating Atrial Fibrillation}
\label{sec:AF}
\begin{figure}[ht]
\includegraphics[width=\textwidth]{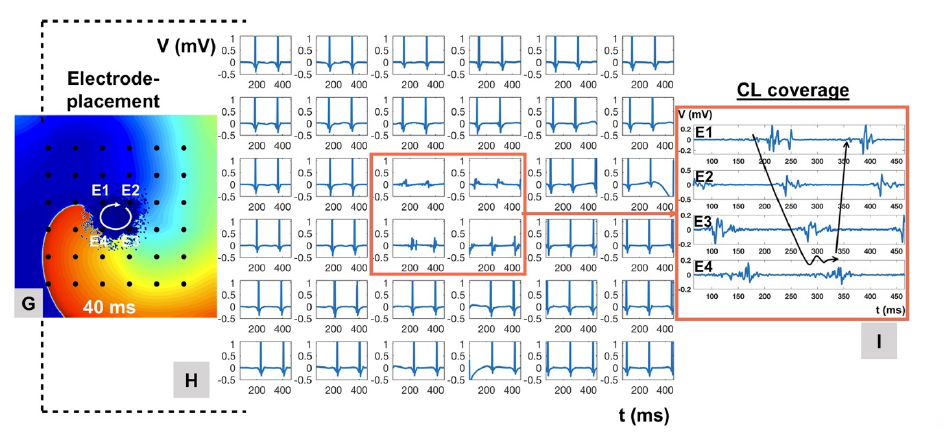}
\caption{A combined clinical and simulation study of electrograms during AFib. For details, see text. Reproduced from Jadidi et al.~\cite{Jadidi-2020-ID14174} under a Creative Commons Attribution 4.0 International License.}
\label{fig:jadidi}       
\end{figure}

Figure~\ref{fig:jadidi} shows the results of a combined clinical and simulation study~\cite{Jadidi-2020-ID14174} investigating several patients with persistent AFib. They frequently found reentry circuits clearly visible with a circular set of electrodes like the Lasso catheter. In the right panel of Figure~\ref{fig:jadidi}, a depolarization wave is moving from E1 via E2 and E3 to E4 and starts again at E1. The signals are fragmented, the voltage is small, and the activity duration is long. To the left, we see the corresponding simulation with fibrotic tissue in the center and very similar signals to the clinically measured ones.

\subsection{Assessing Atrial Arrhythmia Vulnerability}
\label{sec:PEERP}
\begin{figure}[ht]
\includegraphics[width=\textwidth]{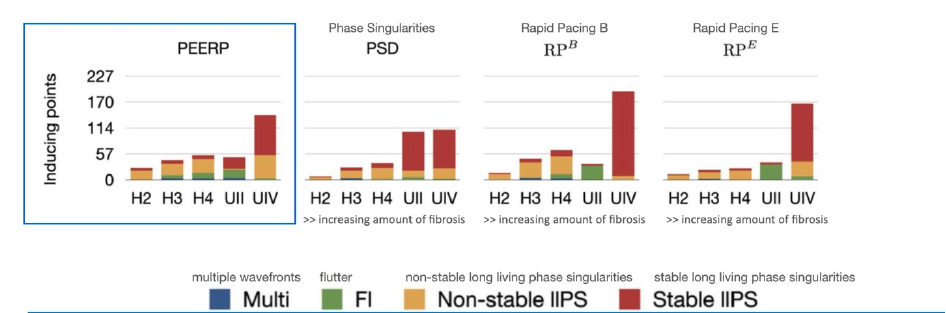}
\caption{Quantitative determination of vulnerability using various methods. For details, see text. Reproduced from Azzolin et al.~\cite{Azzolin-2021-ID16137} under a Creative Commons Attribution 4.0 International License.}
\label{fig:azzolin}       
\end{figure}

A prerequisite for optimizing ablation strategies is a quantitative metric for in silico AFib vulnerability. Otherwise we cannot measure whether one strategy is better than the other. Thus, we need a standardized protocol with trustworthy results to determine the vulnerability of the atria to maintain AFib. We could strictly follow the clinical procedure, which is accelerated pacing at a few points. But we can try to do a better job in silico. Azzolin et al.\ introduced the PEERP protocol: pacing at the end of refractory period~\cite{Azzolin-2021-ID16137}. 
In 7 bi-atrial models, remodeling and various degrees of fibrosis were introduced before attempting to initiate AFib in different ways: with 227 phase singularity points (PS) (about 10\,mm distance), with rapid pacing from 227 points (RP) (about 10\,mm distance) (RP), and with pacing at the end of refractory period (PEERP). The number of inducing points as a quantitative measure of vulnerability and the type of arrhythmia were determined. 
Figure~\ref{fig:azzolin} shows the result: H2, H3, H4 UII and UIV are well defined types and degrees of fibrosis, but here it is only important that the degree of fibrosis is increasing from left to right. The number of points that induced arrhythmias --- the quantitative vulnerability --- are given on the vertical axis. The type of arrhythmia is color coded: multiple wavefronts, AFlut, non-stable long living phase singularities and stable phase singularities. The number of inducing points increased with increasing degree of fibrosis. The amount of phase singularities also increased while the number of multiple wavefronts was small. The number of inducing points was highest when using PEERP, which shows that PEERP is at least as effective in provoking latent arrhythmias as all the other methods. PEERP can uncover vulnerable points that the other methods did not find. The advantage of PEERP: it is much faster than the other methods. Independent of the advantage of PEERP we recognize a major advantage of computerized modeling: it is not possible to test 227 points in the atria for inducing an arrhythmia in the hospital. It takes too much time, and the patient will surely not like it. In computer models one can assess that many points in silico, even before and after introducing interventions. This can help prevent the atria to fall back to AFib much more reliably by also treating latent cases of AFib that did not show up yet in the EP lab, but there is a risk that they will develop in the months after the ablation. In future, surrogate models or machine learning approaches might speed up the process of identifying points vulnerable to stimulation~\cite{MartinezDiaz-2024-ID19750,Bifulco-2023-ID19331,Shade-2020-ID14203}

\section{Personalized Computer Modeling of the Atria}
\label{sec:personalized}
When studying fundamental mechanisms of health and disease, valuable insight can be obtained by using generic models of physiology and pathophysiology. When aiming to support clinical decision making for individual patients directly based on models representing a specific patient, a so-called digital twin, personalization of models is required. For a detailed introduction to the field of digital twins, we refer to dedicated reviews~\cite{Loewe-2022-ID18372,Bhagirath-2024-ID19732,Trayanova-2023-ID18852,Niederer-2021-ID16335}.

Azzolin et al.\ suggested a workflow for atrial model personalization~\cite{Azzolin-2023-ID17490} based on EP and MRI input data – either or, or both of them. First, a personalized geometrical model is created based on the segmented MRI or EP mesh. There is also an option to create the best fitting instance of a shape model~\cite{Nagel-2022-ID17346} to regularize the anatomical model and/or infer missing anatomical structures~\cite{Azzolin-2021-ID16653}. Then, the model is augmented with rule-based anatomical region annotations and fiber direction. Finally, activation times are simulated and compared with the EP data to tune local conduction velocity.  
Other teams published similar approaches with some important differences. Lubrecht et al.\ estimated the local conduction velocities of the left atrium from sparse EP measurements using a monolayer eikonal model~\cite{Lubrecht-2021-ID15889}. Roney et al.\ created bilayer models using EP and MRI data~\cite{Roney-2023-ID18891,Bevis-2024-ID19747} including left and right atrium and defining fibrotic tissue based on the MRI data and assessed inter- and intra-observer variability~\cite{Solis-Lemus-2023-ID18614}. The team of Trayanova used only MRI data for personalization~\cite{Boyle-2019-ID12859}.

Using the approach of Azzolin et al. it is possible to merge EP data with MRI data via the statistical shape model~\cite{Azzolin-2023-ID17490}. This is not as easy as it seems because the geometry from the EP mapping data is most often quite different from the MRI-derived geometry. The statistical shape model is a good method for coregistration. Then, EP data like voltage~\cite{Nairn-2020-ID14911,Nairn-2023-ID18659} and conduction velocity~\cite{Coveney-2022-ID18047,verma18} can be compared with MRI data for identifying fibrotic tissue (IIR value or the Utah value)~\cite{Nairn-2023-ID18842}. 
Figure~\ref{fig:nagel} compares  those modalities based on the individual geometry and a projection of the data to the mean shape~\cite{Nairn-2023-ID18842}. Low voltage areas and slow conduction areas often do not overlap with the fibrotic areas that the two MRI based methods suggest. This is a very puzzling result as computer simulations quantifying AFib vulnerability and ablation success vary depending on which modality was used for modeling fibrosis~\cite{Azzolin-2023-ID17490}. Which data should be used for model personalization?

\begin{figure}[ht]
\includegraphics[width=\textwidth]{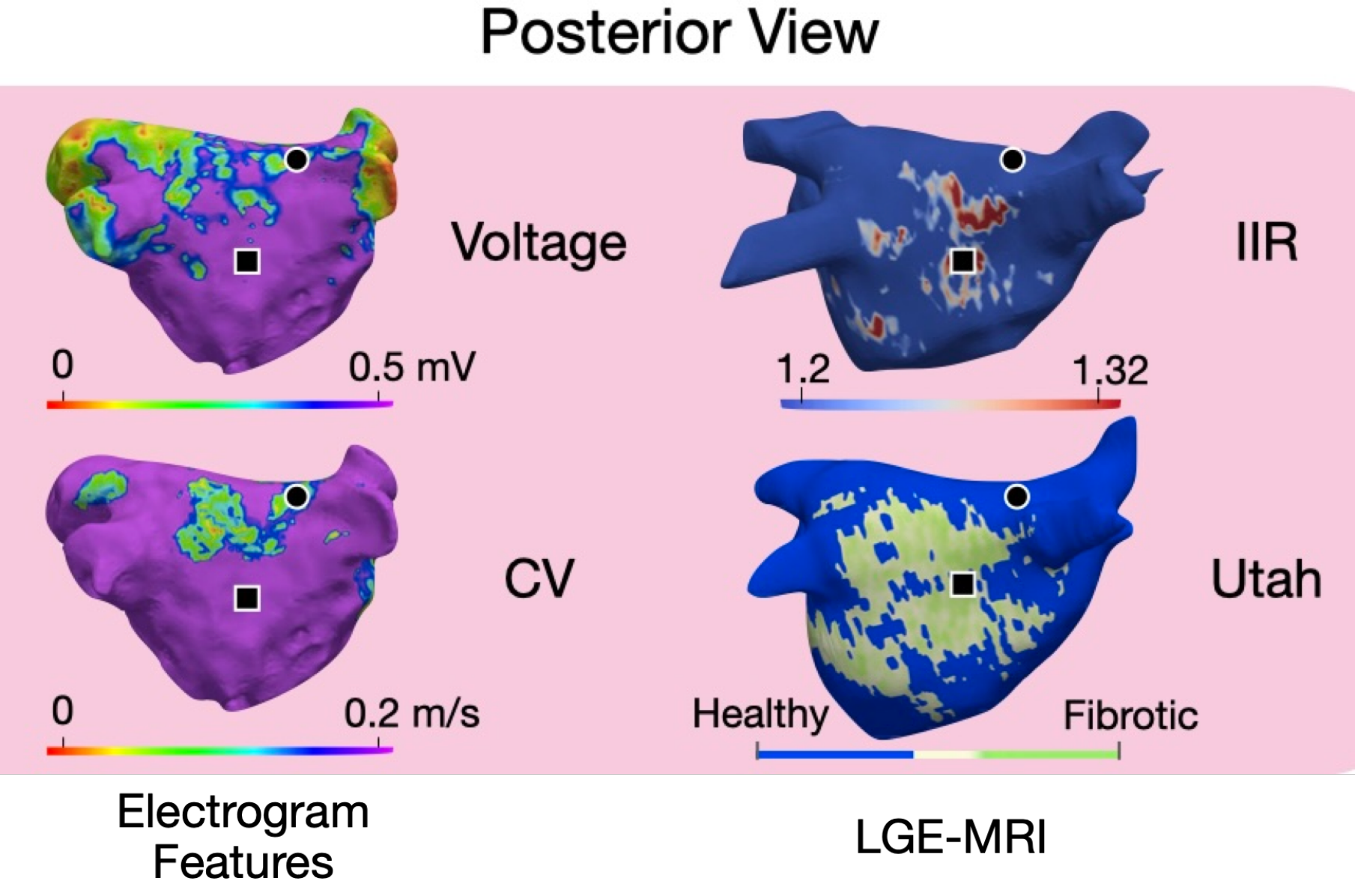}
\caption{Differences in atrial substrate using late gadolinium enhancement-magnetic resonance imaging, electrogram voltage, and conduction velocity. Reproduced from Nairn et al.~\cite{Nairn-2023-ID18842} under a Creative Commons Attribution 4.0 International License.}
\label{fig:nairn}       
\end{figure}

Here is a hypothesis for an explanation of this missing correlation: LGE-MRI is not quantitative due to the so-called partial volume effect. The voxel size of MRI is too large for a quantitative measurement of the degree of fibrosis. There are many voxels in which part of the volume is fibrotic, part is healthy, and part is just blood. It becomes obvious that LGE-MRI data cannot be used for quantitative translation into the degree of fibrosis. But are the data from the electrophysiology lab better? Unfortunately not: in chapter~\ref{sec:egms} we demonstrated that endocardial signals basically see what’s going on on the endocardium and are virtually blind for the epicardial side. It can be argued that fibrosis is only significantly relevant for macroscopic excitation patterns if the degree of fibrosis is between 30\% and 60\%. Below 30\%, it only leads to small changes, and above 60\% it could most simply be considered as a non-excitable area in a computer model. So, all this tells us: LGE-MRI and endocardial EGMs show different aspects of different entities~\cite{Schotten-2024-ID19723}.

At this stage, we suggest to rely more on EP data than LGE-MRI data since AFib is obviously a problem rooted in the electrophysiology of the atria. Nothstein et al.\ demonstrated that the local CV, its anisotropy and its restitution can be measured in a clinical environment using a Lasso catheter~\cite{Nothstein-2021-ID16334}. This is time consuming but including these data e.g. into simulations done with the eikonal model (including the reentry option DREAM) leads to simulation results that nicely fit to clinically measured depolarization patterns (healthy and pathological).

\subsection{Effective Refractory Period}
\label{sec:ERP}
For initiation and maintenance of AFlut and AFib, the effective refractory period (ERP) plays a dominant role. For reentrant activity to be maintained, cells need to have recovered from the previous activation by the time the excitation wave reaches them again. For reentrant activity to be initiated, unidirectional block is one of the dominant mechanisms and favored by local gradients of ERP~\cite{Cluitmans-2023-ID18599}. In a study by Martínez Díaz et al., clinical ERP measurements were obtained in 7 patients from multiple locations in the atria~\cite{MartinezDiaz-2024-ID19469}. Atrial computer models with personalized and non-personalized ERP data were compared. Inducibility of AFib was higher in personalized scenarios compared to scenarios with uniformly reduced ERP. Unger et al. investigated a more practical approach to estimate the ERP from clinical data~\cite{Unger-2021-ID15892}.

\section{Modeling Cohorts of Patients}
\label{sec:cohorts}
The objective of modeling cohorts of patients should be clearly separated from the former chapter about personalization – although the two aims are obviously linked. Modeling cohorts aims to represent real-life (sub-)populations in statistical terms and can allow for clustering of patients. It may happen that a fully personalized model of the atria is nice to have but not practical, in particular if we think of the millions of patients suffering from AFib. A practical approach could aim at defining for example 5 classes of AFib patients and identify the class an individual patient belongs to based on a few simple and cheap measurements. For each class, standardized ablation strategies that will presumably be most successful can then be identified. Modeling of cohorts can support this approach. 

Another option is using cohorts of patients for setting up a training data set for machine learning~\cite{Loewe-2022-ID18371,Alber-2019-ID19733}. Gillette, Gsell, Nagel et al.\ modeled a cohort of nearly 17,000 virtual patients to simulate a comprehensive and fully traceable ECG database~\cite{Gillette-2023-ID18724}. For the cohort of atrial models and corresponding P-waves in the ECG, 80 biatrial geometries from a statistical shape model were used. These models were augmented with several types and degrees of fibrosis, several torso geometries and several positions of the heart inside the thorax. The P-waves that were simulated match most of the P-wave feature distributions of large repositories of measured ECGs providing evidence for the credibility and realism of these simulations.

\begin{figure}[ht]
\includegraphics[width=\textwidth]{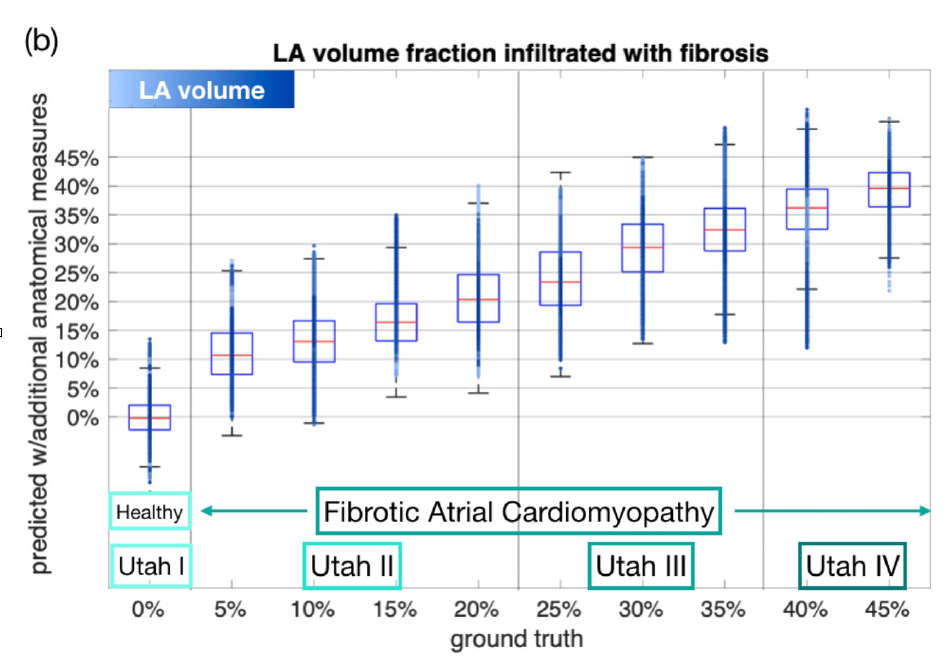}
\caption{Non-Invasive and quantitative estimation of left atrial fibrosis based on P-waves of the 12-Lead ECG. Reproduced from Nagel et al.~\cite{Nagel-2021-ID16114} under a Creative Commons Attribution 4.0 International License.}
\label{fig:nagel}       
\end{figure}

Nagel et al.\ also simulated even more than 500,000 ECGs with the objective to estimate the degree of fibrosis in patients just from the 12-lead ECG using machine learning for AFib risk prediction~\cite{Nagel-2021-ID16114}. A neural network classifier was trained using the simulated data. Figure~\ref{fig:nagel} shows the result: On the x-axis, the ground truth is depicted, i.e. the real degree of fibrosis as introduced in the model and on the y-axis the predicted degree of fibrosis is shown  – just based on the P-wave. There is a considerable quantitative error but in most cases the correct Utah class is identified, which is used clinically for patient stratification and which is based today on expensive LGE-MRI exams and might be replaced by using the ECG only. In a followup study, this approach could be validated on an independent test set in comparison to low-voltage-based identification of fibrotic substrate~\cite{Nagel-2022-ID18491}.

\section{Modeling Ablation Strategies}
\label{sec:ablation}
Which ablation strategy is optimal for an individual patient? Can we not only stop the current AFib episode but also prevent future episodes? Several articles have been published about this topic~\cite{bayer16, Boyle-2019-ID12859, Shim-2017-ID11782, Roney-2020-ID14699, Alessandrini-2018-ID12168}.
Azzolin et al.~\cite{Azzolin-2022-ID18076} implemented a virtual environment in which ablation lines can be applied in virtual personalized atria of a patient to determine immediately what happens to the vulnerability with respect to AFib. This allows for a systematic comparison of standard strategies like isolation of the pulmonary veins, anatomical, and substrate-guided ablation. The clear advantage compared to a clinical situation: if you do not like the ablation lines any more you press the reset button and you can start all over again. This `learning by burning approach' is obviously not possible with real ablation lines.

Azzolin assessed 13 ablation strategies and determined the vulnerability for AFib reccurrence after the ablation procedure in a cohort of 29 patients with various degrees and types of fibrosis. The subset of substrate-based strategies were informed by low voltage in EGMs, low conduction velocity, high dominant frequency, and LGE-MRI. Most of them were combined with pulmonary vein isolation. The result: ablation of areas with high dominant frequencies in silico in combination with connecting ablation lines to the nearest orifice were most successful. Omitting these connections led to $\approx$20\% lower success rates. Also imprecise localization of the fibrotic tissue (LGE-MRI vs. low voltage based) led to $\approx$35\% lower success rates~\cite{Azzolin-2022-ID18076}.

\section{Summary}
\label{sec:summary}
Realistic and useful computerized modeling of atrial electrophysiology and AFib is more complicated than we originally thought. We can model the healthy atria with high fidelity. But to be useful for patients, we need to model diseased tissue and arrhythmias. For that purpose, we need more data about the underlying disease(s). We need the right model for the substrate of AFib: type and degree of fibrosis, myopathy, scar. We need better knowledge about which mechanisms can initiate atrial arrhythmias and which kind of substrate will perpetuate atrial arrhythmias. Can we reliably estimate the vulnerability of an atrium to develop an arrhythmia using standardized and reproducible in silico protocols? 

We need to answer a number of questions: Which data do we really need for useful personalization? Can we use rule-based techniques and rely on population-level information? Can we measure these data in a clinical setting with sufficient accuracy? Which parameters of our model vary over time and in space and how large is this variation? 

In this article we have shown results that tell us: for reliable ablation planning, we need a personalized 3D model or a bilayer model of the atria, we need fiber direction (rule-based) and anisotropic CV, we must include realistic assumptions/estimates about remodeling, CV restitution and regional effective refractory period. Extended eikonal models could be a good compromise for obtaining trustworthy simulation results within short calculation times. Using a detailed model of fibrosis, we could extract important information about the type and degree of fibrosis based on fractionation in high-quality electrograms.

Computerized models are a mighty tool for research and potentially also for better diagnosis and treatment of patients. However, these models must build on experimental data that have to be determined in many patients to be statistically valid in a wide and diverse range of patients. 
And we should not forget: before these computational models can move into clinical use, they must pass clinical validation demonstrating that this model really fulfills the promise that is made, e.g. improving patient outcome in ablation of atrial arrhythmias. 

\bibliographystyle{spmpsci}
\bibliography{references}
\end{document}